\documentclass[9pt]{article}

\usepackage{authblk}
\usepackage{amsmath}
\usepackage{graphicx}
\usepackage{fullpage}
\usepackage{hyperref}

\begin{document}

\title{Saccade crossing avoidance as a visual search strategy}

\author[1,*]{Alex Szorkovszky}
\author[2,3]{Rujeena Mathema}
\author[2,3]{Pedro Lencastre}
\author[1,2,3,4]{Pedro G. Lind}
\author[2,3,5]{Anis Yazidi}

\affil[1]{Department of Numerical Analysis and Scientific Computing, Simula Research Laboratory, Oslo, Norway}
\affil[2]{Department of Computer Science, Oslo Metropolitan University, Oslo, Norway}
\affil[3]{OsloMet Artificial Intelligence Lab, Oslo Metropolitan University, Oslo, Norway}
\affil[4]{School of Economics, Innovation and Technology, Kristiania University of Applied Sciences, Oslo, Norway}
\affil[5]{Department of Informatics, University of Oslo, Oslo, Norway}
\affil[*]{E-mail: alex@simula.no}

\date{}

\maketitle

\begin{abstract}
Although visual search appears largely random, several oculomotor biases exist such that the likelihoods of saccade directions and lengths depend on the previous scanpath. Compared to the most recent fixations, the impact of the longer path history is more difficult to quantify. Using the step-selection framework commonly used in movement ecology, and analyzing data from 45-second viewings of {\em Where's Waldo?}, we report a new memory-dependent effect that also varies significantly between individuals, which we term {\em saccade crossing avoidance}. This is a tendency for saccades to avoid crossing those earlier in the scanpath, and is most evident when both have small amplitudes. We show this by comparing real data to synthetic data generated from a memoryless approximation of the spatial statistics. Maximum likelihood fitting indicates that this effect is strongest when including the last $\approx 7$ seconds of a scanpath. The effect size is comparable to well-known forms of history dependence such as inhibition of return. A parametric probabilistic model including a self-crossing penalty term was able to reproduce joint statistics of saccade lengths and self-crossings. We also quantified individual strategic differences, and their consistency over the six images viewed per participant, using mixed-effect regressions. Participants with a greater tendency to avoid crossings displayed smaller saccade lengths and shorter fixation durations on average, but did not display more horizontal, vertical, forward or reverse saccades. Together, these results indicate that the avoidance of crossings is a local orienting strategy that facilitates and complements inhibition of return, and hence exploration of visual scenes.
\end{abstract}

\vspace{0.5cm}

Statistically speaking, visual scan paths when viewing scenes look remarkably like random walks, particularly when salient areas are evenly distributed \cite{boccignone2004modelling,schutt2017likelihood}.
Comparisons have frequently been made to animal foraging \cite{bella2022foraging} --- although in the case of vision, movement occurs primarily in discrete steps (saccades), between which are relatively still periods (fixations). Yet, as with animal foraging, there are task-dependent biases and patterns that can be uncovered if enough eye-tracking data is recorded \cite{tatler2009prominence}, and which can increase time and/or memory efficiency of search \cite{najemnik2005optimal}. Quantifying these biases and their correlations is useful for understanding neural pathways involved in attention and search \cite{wagner2024individual} and can also help to diagnose neurological conditions \cite{lagun2011detecting,archibald2013visual}.

Some biases, such as a tendency for the eyes to relax towards the centre of vision, can be specified as a simple function of the most recent fixation (i.e.\ fixations from this point toward the centre become more likely than in the opposite direction). In general, a Markovian model such as $P(x_{t+1},y_{t+1}|x_t,y_t)$ \cite{clarke2017saccadic}, when combined with saliency modelling, can come close to deep learning models in predicting the next fixation \cite{kummerer2021state}. Such a distribution is also low-dimensional enough to generate with nonparametric methods \cite{lencastre2023modern}.

Other biases depend on the most recent saccade (i.e.\ the two most recent fixations), such as the increased likelihood of forward and return saccades, which continue with the same length and in the same or reverse direction, respectively \cite{wilming2013saccadic}. In other directions, regular changes in saccade length are more common than what is expected if subsequent lengths are assumed to be independent, leading to a negative autocorrelation \cite{malem2020mathematical}.

Finally, some biases have longer history-dependence. The probability of saccades returning to `$k$-back' fixations decreases only slowly with $k$ in some tasks \cite{wilming2013saccadic}. This is has been called ``facilitation of return'' \cite{smith2009facilitation}, in competition with ``inhibition of return'' which supposedly facilitates exploration \cite{klein1999inhibition}. It has been proposed that the former bias applies for salient points which require further consideration \cite{wilming2013saccadic}, while the latter is more prominent for simple distractors and low working memory demands \cite{shen2014working}.

The aforementioned influences on saccade dynamics have been quantified and modelled in various ways. Often they are based on comparing distributions of saccade and fixation measures to models assuming the absence and/or presence of some effect. The distributions compared (such as saccade length or change in saccade angles) depends on the phenomenon being examined. While there have been recent attempts to unify modeling approaches under the banner of likelihood-based inference \cite{kummerer2015information}, this has mainly been applied to comparing overall models against a benchmark rather than comparing the importance of predictors within models.

Another difficulty with identifying history dependence is the short length of most tasks and datasets. Deep learning models have been used to show that up to four fixations can increase predictive power in free viewing \cite{kummerer2022deepgaze}. However, given what is known about refixations, it is expected that humans should be able to keep track of a longer scanpath during search.
\begin{figure*}[t]
\centering
\includegraphics[width=0.8\linewidth]{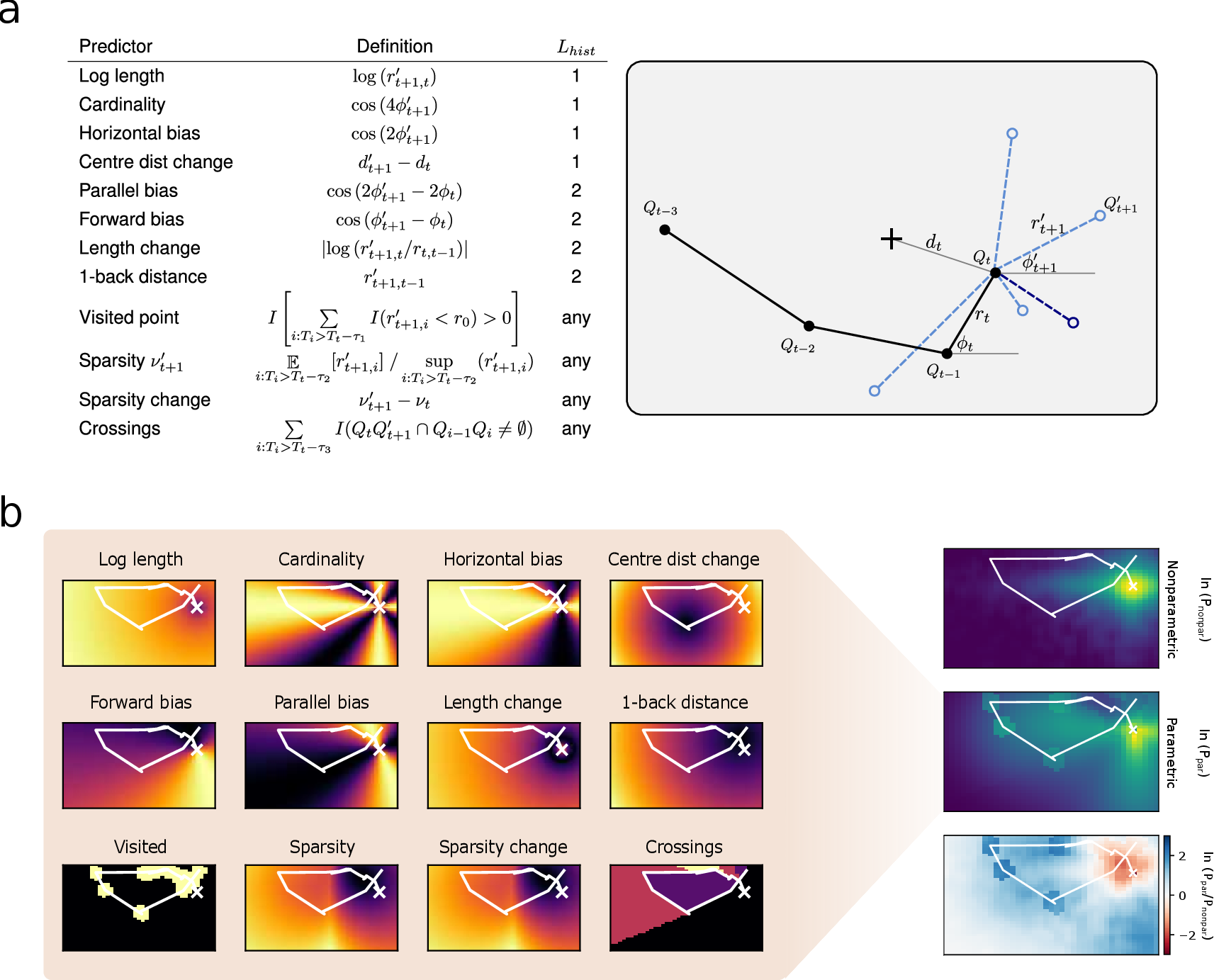}
\caption{(a) List of predictors used (left), sorted by the number of most recent fixations $L_{hist}$ required to calculate them; and an example of a scanpath illustrating key quantities and case-control sampling (right). Saccades are approximated as straight lines between mean fixation locations. In the illustration, predictors are calculated for fixation $Q_{t+1}$ by generating four random control saccades (light dotted lines) in addition to the saccade from the data (dark dotted line). The center of the screen is indicated by a cross. 
(b) The panels on the left illustrate parametric predictor values for test points $Q'_{t+1}$ given a scanpath up to $Q_t$ indicated by an X. The panels on the right show overall spatial log likelihoods for the nonparameteric model (top) the combined parameteric model (middle) and the difference (bottom). 
}
\label{fig:predictors}
\end{figure*}

Here, we use step selection analysis, a case-control sampling based method common in studies of animal movement \cite{forester2009accounting,thurfjell2014applications,michelot2024understanding}, to quantify biases and long-term history dependence in scanpaths during visual search. There are several advantages of this approach: Firstly, by using a typical saccade model to sample both taken and available steps, it can account for the fact that no two data points will have the same path history (in analogy with the heterogeneity of environmental features encountered by animals as they move). Secondly, being compatible with general regression models, it can evaluate the relative importance of and interactions between known biases in a straightforward way. Thirdly, by including random effects, it can also capture repeatable variation in biases between individuals \cite{muff2020accounting}, which would otherwise require a large amount of data per individual to quantify. Previous studies of strategic variation have focused on simple measures such as saccade frequency \cite{boot2009stable}. 
Given that visual search performance is correlated with several personality traits and cognitive abilities \cite{wagner2024individual}, we expect the underlying strategy to vary consistently along several other measures. While consistent differences have been identified in saccade lengths and fixation durations across tasks and time \cite{andrews1999idiosyncratic,rayner2007eye,henderson2014stable}, stable patterns involving several measures have not yet been identified in visual search.

In addition to the most commonly reported biases and memory-dependent effects, we propose a new bias that we term {\em self-crossing avoidance}. This is inspired by the concept of space-filling curves, which exhibit this feature as a means to efficiently cover an area. The concept of self-avoiding random walks has also been explored by Engbert et al. to explain the statistics of small-scale eye movements during fixations \cite{engbert2011integrated}, although in this case self-avoidance emerges from a combination of inhibition and a uniform microsaccade length. Instead, we include the number of recent saccades that are crossed by a candidate saccade as a potential influence on its likelihood.

We use the book series {\em Where's Waldo?}, which has been used for several other studies of visual search \cite{klein1999inhibition,dickinson2005marking,smith2011looking,zhang2018finding}. With a high density of salient and heterogeneous distractors, this is a relatively difficult search task compared to natural scenes \cite{zhang2018finding} and categorical target-distractor paradigms \cite{wolfe2021guided}. It is known that the density of visual features influence search patterns \cite{wolfe2021guided}. However, the relative homogeneity of the Waldo paradigm presents a task where the search space is not biased by more and less salient regions. Therefore, this is highly suitable for ``pure'' image-independent scan-path modelling, as well as for uncovering variation in search strategies.

To compare the search task to a bottom-up free viewing, we also use stimuli consisting of random coloured pixels, in which the participant is instructed to look for shapes and patterns hidden in the image. This presents another homogeneous stimulus for identifying differences between individual scan patterns \cite{judd2011fixations}.

\section*{Results}

The predictors used are defined in Figure \ref{fig:predictors} and visualized for a sample scanpath.
The predictors that depend on more than one previous saccade each contain a time constant that truncates the history length. These were optimized first separately and then jointly to maximize the likelihood of the logistic regression model (see Supplementary Information). For the Waldo task, the binary {\em visited point} predictor, indicating whether or not the candidate point was within 1.5 degrees (approximately the foveal angle) of a recent fixation, was optimal at $\tau_1=6$ s. The {\em sparsity} predictor, quantifying the ratio of the mean to maximum distance of previously visited points, was strongest when $\tau_2$ was set to 32 seconds. Finally, the {\em self-crossing} predictor, indicating the number of times a candidate saccade crossed the recent scanpath (determined algebraically from the intersection of straight line segments) was optimal at $\tau_3=7$ seconds. When optimized jointly, $\tau_1$, $\tau_2$ and $\tau_3$ moved to 4, 33 and 6 seconds respectively. The three features therefore appear to capture complementary aspects of the path dependence, as shown by the lack of strong dependence between the time constants (see Supplementary Figure S1).

\begin{figure}[t]
\centering
\includegraphics[width=0.5\linewidth]{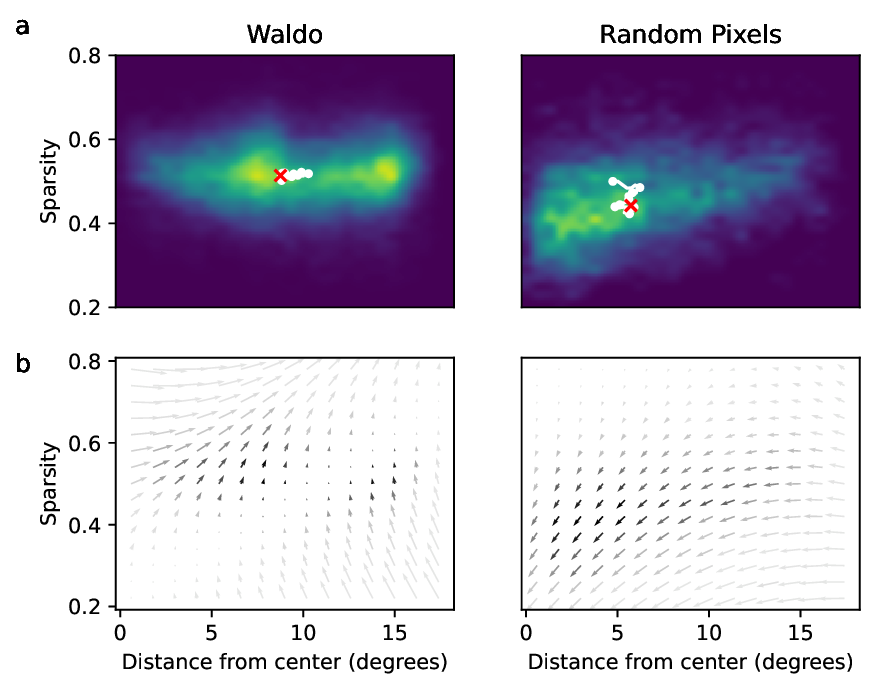}\\
\includegraphics[width=0.5\linewidth]{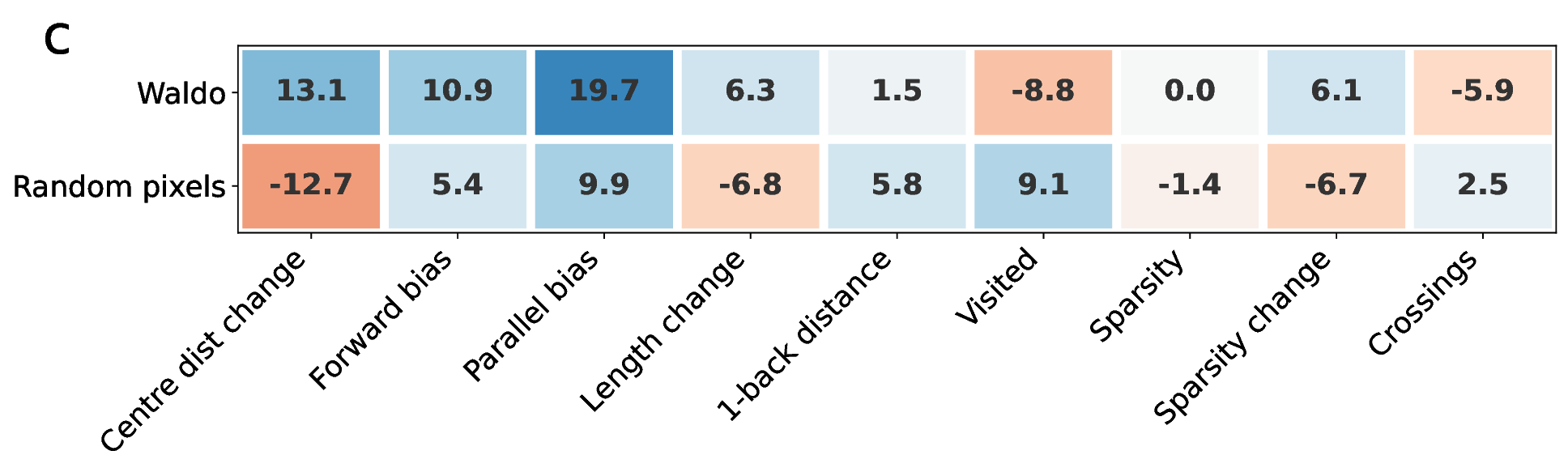}
\caption{Task comparison. The panels in (a) compare the spatial distributions of fixations in terms of sparsity and distance from the centre of the image. Overlaid are moving averages over time, with a window length of 5 seconds, ending in the red crosses. The arrows in (b) shows the direction and slope of maximum gradient in log-odds from fitted parametric models evaluated at the small-saccade limit $Q'_{t+1}\approx Q_{t}$ (see Supplementary Information). The half-width of the image is approximately 15.2 degrees. Panel (c) compares normalized z-scores for the main predictors in the model. The color of each cell gives the z-score when including only a linear term for this predictor and all interactions for all other terms. Terms that do not decrease the AIC are given a z-score 
of zero. Log-length, horizontal bias and cardinality are not included due to strong dependence on the saccade model.}
\label{fig:spatial}
\end{figure}

The sparsity measure, along with the distance from the center, partially capture how fixations were distributed overall within the screen area and among previous fixations. As shown in Figure \ref{fig:spatial}(a-b), these distributions differ strongly between the the two tasks. Panel (b) shows that for the Waldo task, the likelihood of shifts in these quantities were more dependent on their current values; while (c) shows that, overall, changes in these predictors had opposite effects on fixation likelihood.

As expected, the history-independent predictors were the strongest overall, as shown in Figure \ref{fig:spatial}(c) and Supplementary Figure S4. {\em Parallel bias}, which equally captures forward and return saccades, was the strongest linear predictor in the Waldo dataset. This, the additional {\em forward bias}, and one-back distance were the only predictors that were in the same direction for the random pixel data.
The predictors related to previously visited points indicated an inhibition of return on average in the Waldo task, as shown by the positive effect of change in sparsity and negative effect of a candidate fixation being close to one in its recent history, and conversely a facilitation of return for the random pixel task.

\begin{figure}[t]
\centering
\includegraphics[width=.5\linewidth]{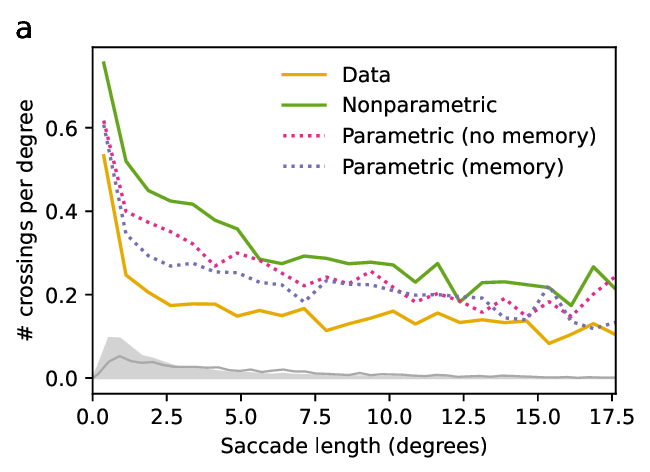}\\
\includegraphics[width=.5\linewidth]{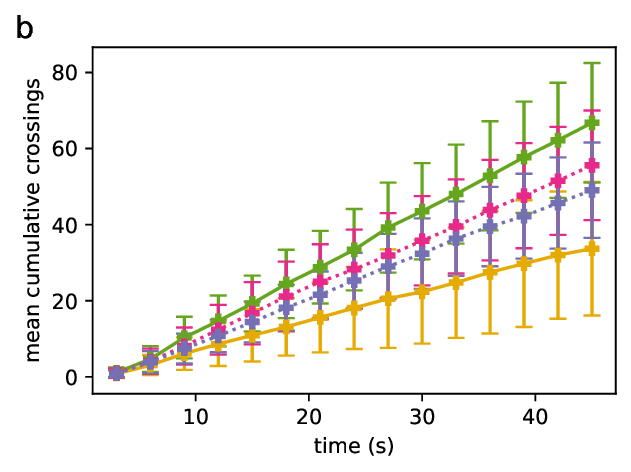}
\caption{(a) Mean number of crossings (including intersections with both past and future saccades within $\tau_3$) divided by saccade length, plotted against saccade length, for the models and test data. The grey area shows the density of saccade lengths in the test data, while the grey line shows the density of saccades with crossings.
(b) Cumulative number of crossings over time for data generated from each model along with the test data. Shown are means and standard deviations over all 70 trials in each set.} 
\label{fig:genmodel}
\end{figure}

For the Waldo task, synthetic data was generated to further investigate the self-crossing statistics. A nonparametric and memoryless null model was fit, based on a kernel-smoothed empirical distribution $P(\Delta x_t,\Delta y_t|x_t,y_t)$. To control for the well-known coarse-to-fine transition \cite{over2007coarse}, in which mean saccade length decreases over time, a cubic spline was fit to the saccade log-length as a function of time. Generated saccades were then rescaled according to this curve. The resulting generative model matched the overall distribution of saccade lengths for the entire trial length (see Figure \ref{fig:genmodel}(a) and Supplementary Figure S3). For comparison, the same amount of data was also generated using the parametric model fit by the step selection process.

For disconnected and independent line segments, crossing probability is expected to increase proportionally to length \cite{cowan1989objects}. Here, since fixations are spatially clustered, there is a decreased probability of crossing per unit length for longer saccades (see Figure \ref{fig:genmodel}a). Compared to the null model, the experimental data showed relatively fewer crossings for saccades of all lengths, particularly those of intermediate length. The experimental trend was best approximated by the memory-dependent parametric model. This was also true for ``crossing'' and ``crossed'' saccades separately (see Supplementary Figure S5), despite existing saccade lengths not being a model input.
Overall, the mean number of self-crossings per saccade was $0.49$ compared to $0.25$. As shown in Figure \ref{fig:genmodel}b, the full parameteric model was closer to the self-crossing distribution of the data than to the null model, with a mean of $0.36$ crossings per saccade. 


\begin{figure}[t]
\centering
\includegraphics[width=.5\linewidth]{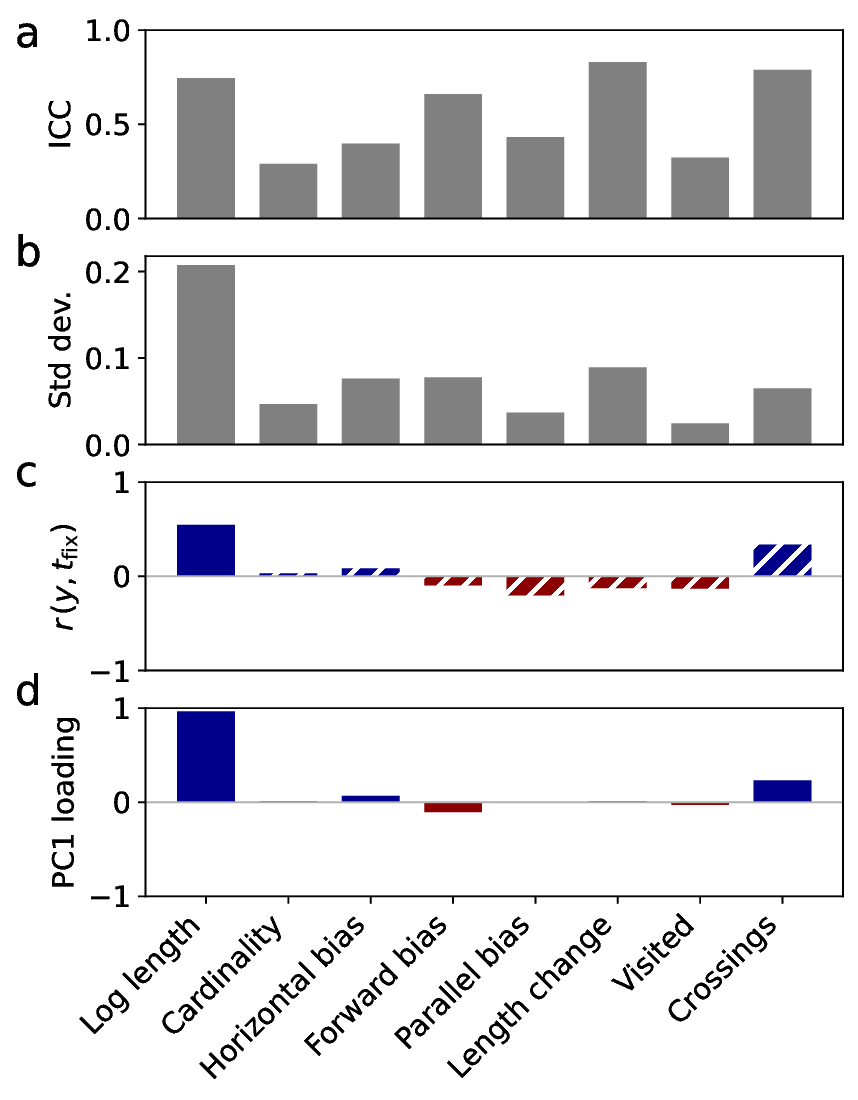}
\caption{Waldo task individual differences. (a) Intraclass correlation coefficients for predictors where there was significant variation among both trials and participants. (b) Corresponding standard deviations of the participant-level random effect. (c) Pearson correlation coefficients between each significant random participant effect $(y)$ and the mean of the trial-level median fixation duration $t_\mathrm{fix}$. Hatched bars indicate lack of significance at the $p<0.05$ level. (d) Loadings of the first principal component given the fitted means for each participant and each significant predictor.}
\label{fig:variation}
\end{figure}

Random effects were then added sequentially to each feature of a linear step selection model to quantify the individual variation in search strategies. For the Waldo task, all features had a significant participant effect (according to a reduced AIC) apart from {\em center distance change}, {\em 1-back distance}, {\em fixation sparsity} and {\em fixation sparsity change}. The intraclass-correlation coefficient, quantifying how much of the variation across trials was due to inter-participant variation, was highest for length change and self-crossing variance (ICC $0.74$ and $0.79$ respectively, see Figure \ref{fig:variation}(a)). 

Several predictors were also positively correlated with those fitted for the random pixel task (without trial effects in the latter models due to the low number of trials per participant) at the $p < 0.05$ level, with only the {\em cardinality} and {\em visited} predictors not being significantly correlated across the $n=31$ participants. The remaining Pearson correlations ranged from $r=0.38$ ($p=0.033$) for self-crossing avoidance to $r=0.61$ ($p<0.001$) for forward bias.

The significant participant-level random effect for log length was positively correlated with average fixation durations in the same task ($r=0.55, p=0.001$, see Figure \ref{fig:variation}c, Supplementary Figure S8). When averaged over all participants, fixation time was greater for the random pixel task (mean 0.40 vs 0.27 for Waldo, t-test $t=10.3, p<0.001$). Despite this large overall difference, the individual-level average fixation times were positively correlated between the two tasks (Pearson $r=0.43,p=0.016$).

When dimensionality of the unscaled random effects was reduced using principal component analysis, the first component captured 65\% of the variance for the Waldo task. Participants that scored high on this component took longer steps, with more self-crossings (see Figure \ref{fig:variation}d). As was the case for these predictors on their own, the first principal component was also correlated with mean fixation duration ($r=0.54, p=0.002, n=31$). The second principal component, with 18\% variance explained, was highest for participants with more forward bias, more horizontal bias, and less alternation in saccade length (see Supplementary Figure S9). The PCA loadings for the random pixel task were markedly different, and were not correlated with fixation time. However, the random effect for change in centre distance was negatively correlated with mean fixation time in this case ($r=-0.37$, $p=0.04$, see Supplementary Figure S6).

\section*{Discussion}

Difficult visual search tasks can take a considerable amount of time, providing ample opportunities to investigate participants' search strategies as well as cognitive and oculomotor biases. The latter are more often studied using short-length free-viewing datasets designed for studying saliency, since these are much more widely available \cite{kummerer2021state}. This means that long-term effects are generally under-appreciated when studying scanpaths.

Using the step selection framework on long scanpaths allowed various influences on fixation selection to be considered simultaneously and to be compared on an equal footing. This technique is therefore suitable for tasks such as free viewing or visual search of scenes, and lessens the need for exhaustive experimental manipulations which, while often cleverly designed \cite{ehinger2018probing}, ultimately lessen ecological validity. Although we did not use image features, step selection can easily integrate aspects of image saliency, which can be used to compare top-down and bottom-up influences on the scanpath.

People looking at random, information-poor images will tend to focus more on the center of the image \cite{judd2011fixations}. Comparing step selection results between the Waldo and random pixel tasks shows that not only is this the case, but the image center is attractive on average for random images and repulsive for search. Opposite directions of influence were found for several memory-dependent and memory-free biases.

Since the resolution of spatial information in memory decreases over time \cite{sheth2001compression,dickinson2007memory}, the longest term predictors should be the ones that require the least spatial resolution. Consistent with this, our fixation sparsity measure is insensitive to small changes in the scanpath history and had the longest optimal memory length. Likewise, the visited point predictor requires resolution at the level of the foveal area and had the shortest optimal memory length. 

We introduced the number of saccade self-crossings as a useful characteristic of scanpaths, which was reduced in the search task compared to a memoryless model. A full generative model approached the empirical data, partially due to explicit self-crossing avoidance. 
The self-crossing tendency showed strong inter-individual variation, which could mean that some participants pursued a systematic strategy of organized scanpaths, as shown in \cite{gilchrist2006evidence} for stimuli on a grid. However, this would also imply strong variation in and correlation with cardinality, which was not seen in our data. We found that saccades of all lengths tended to be avoided compared to the null model. This indicates a need for high spatial resolution in memory, consistent with the shorter optimal history length for this predictor.

Another clue for the functional significance of this variation can be found in the duration of fixations, widely used as a proxy for focused visual processing \cite{nuthmann2017fixation}. These were positively correlated at the individual level with saccade length and self-crossings in the Waldo task, which comprise the primary axis of variation in this task. Unlike the commonly found negative relationship between saccade length and fixation duration, as seen when comparing ambient and focal viewing over time \cite{unema2005time}, or between novices and experts \cite{sheridan2017holistic}, this relationship was instead found to be positive at the inter-individual level for the Waldo task. Such a relationship at the inter-individual level has been reported for free viewing \cite{risko2012curious}, but previous studies for search, such as Henderson and Luke \cite{henderson2014stable}, may have had too small a sample size to reach statistical significance. 

Correlation between fixation duration, saccade length and self-crossing avoidance may indicate that the latter is used primarily during the faster process of spatial orientation \cite{trevarthen1968two}, and some participants use more of this ``ambient'' processing. In aviation, for example, experts use more complex scanpaths and shorter fixations than amateurs \cite{lounis2021visual}. Together, this may indicate that avoiding crossing of recent saccades is a relatively simple orienting strategy that facilitates exploration and complements explicit inhibition of return to recently visited locations.
How this strategy depends on attentional demands could be studied, for example, with multiple-target search using a variable number of features for target matching \cite{kristjansson2014common}. Also, the possibility of long-term forward planning could be investigated by testing whether self-crossing remains in a gaze-contingent paradigm \cite{dickinson2005marking}.

This work also has implications for deep learning scanpath models. The state-of-the-art Deepgaze III model \cite{kummerer2022deepgaze} uses three images per fixation to encode their positions, so that they can be input into a convolutional neural network (CNN) and combined with a saliency map. This means that the size of the scanpath network will be proportional to the desired history length. Therefore, for deep learning to be useful in modeling human attention during search, it may be necessary to find more compact representations of history. Heatmaps of our three long time-scale heuristics, which we included for illustration, may also be useful as CNN inputs in this respect. 

We have demonstrated a new and powerful method for quantifying oculomotor biases, leveraging techniques used in movement ecology. Nonlinear methods such as generalized additive models or neural networks may increase the match to the data, however this would come at the expense of easy interpretability that the logistic regression provides. We expect that incorporating image features may provide both competitive and insightful results on common saliency benchmarks. Future work that correlates strategic variation with cognitive and personality differences may also shed light on the ways neural circuitry is recruited during visual search.

\section*{Materials and Methods}
\subsection*{Data}
31 participants, using chin rests approximately 1 meter from a 1920x1080 pixel monitor, were each shown nine {\em Where's Waldo} images that filled the screen for 45 seconds each, followed by three images comprised of randomized RGB values per pixel, for 60 seconds each. Eye gaze was tracked using the EyeLink Duo at 1000Hz frame rate. Saccades and fixations were automatically extracted using the default settings (30 $^\circ$/s velocity, 8000 $^\circ$/sec$^2$ acceleration) which do not detect microsaccades. Fixations outside the image boundary were discarded. One in every four trials were reserved as a test set, distributed evenly among individuals and among images, with the remaining used for training the generative models. All data was used for the final fitting of mixed effect models. Data and analysis code are available online\footnote{\url{https://doi.org/10.5281/zenodo.17378732}}.

\subsection*{Step selection}
Importance sampling was used for generating control points \cite{forester2009accounting,michelot2024understanding}. For each real data point, nine control data points were generated from a radially symmetric lognormal distribution fitted to the training set, with sampling repeated for any points generated outside the image boundary. For each data point, all predictors were calculated, along with the square of the log-length. These were then fed into a logistic regression to separate data points from control points. All interaction terms were initially included, and stepwise selection was used to remove insignificant interaction terms (see Supplementary Information). An interaction term between log length and time was also included to capture the coarse-to-fine transition \cite{over2007coarse}. Lengths are measured in pixels for the regression models. These regressions were done using the Python {\em statsmodels} package.

\subsection*{Time constant optimization}
For each variable-history predictor, the cutoff $\tau$ was iterated in units of one second up to the trial length of 45 seconds (see Supplementary Information). The value of $\tau$ maximizing the likelihood of the step selection, when including this predictor, the short-term predictors and all interactions, was then obtained and used for the rest of the analysis. For fixation sparsity, both the sparsity and the change in sparsity were included with the same value of $\tau_2$.

\subsection*{Nonparametric generative model}
Gaussian kernel density estimation was used to smooth the conditional empirical distribution of $P(\Delta x_i,\Delta y_i|x_i,y_i)$. One variance was used for $x,y$ and another was used for $\Delta_x,\Delta,y$, and these were optimized jointly (see Supplementary Information). This was sampled from directly to generate synthetic data. To account for the change in average saccade length over time, a cubic spline was fit to the empirical log-length against time. Saccade lengths were then adjusted based on the time in the simulation according to the spline. Any points outside the boundary of the image were discarded and resampled. Time intervals between fixations were sampled from its unconditional distribution, which determined the total number of fixations within the trial length.

\subsection*{Parametric generative model}
At each time step, 40 candidate points were generated as for the step selection process, resampling any outside the image boundary. The fitted step selection model was then used to estimate the likelihood of each point, adjusting the intercept term for the difference in number of samples. One sample was then chosen with a probability proportional to this likelihood. Time intervals were sampled from the unconditional distribution as for the nonparametric model.

\subsection*{Mixed-effect modelling}
Individual variation was tested and quantified using a binomial mixed effect regression. All predictors were first centered and normalized. In the Waldo task, for each of the fixed effects, a term modeling the slope varying over trials (279, nine per participant, variance $\sigma_t^2$) was tested first against the base model with only fixed effects using the AIC. If the mixed model had a lower AIC, a second random slope effect was included for participant (31, variance $\sigma_p^2$), and also tested using the AIC. For predictors with a significant participant effect, the ratio $\sigma_p^2 / (\sigma_p^2 + \sigma_t^2)$ was used as the intraclass correlation coefficient (ICC). For the random pixel task, only one test was performed for the participant effect, due to the small number of trials, and thus the ICC was not obtained. The {\em R} package {\em lme4} was used via the Python package {\em pymer4}\cite{jolly2018pymer4}.

\section*{Acknowledgements}
This work is funded by the Research Council of Norway's FRIPRO grant number 335940. The authors would like to thank Gustavo Mello and Jer\^{o}me Tagu for helpful discussions.

\bibliographystyle{unsrt}
\bibliography{selfcrossing}

\end{document}


\title{Saccade crossing avoidance as a visual search strategy:\\Supplementary Information}
\author{Alex Szorkovszky\thanks{Corresponding Author: alex@simula.no}, Rujeena Mathema, Pedro Lencastre, Pedro G. Lind, and Anis Yazidi}

\date{}
\maketitle

\setcounter{figure}{0} \renewcommand{\figurename}{Fig.} \renewcommand{\thefigure}{S\arabic{figure}}



\section{Eye tracking}

Calibration was performed using a five point method. The EyeLink's internal classifier was used to classify saccades, fixations and blinks. The default threshold values were used for saccade detection (30 $^\circ$/s velocity, 8000 $^\circ$/sec$^2$ acceleration). The right eye positions and classifications were used for further analysis.

\section{Preprocessing}

After removing blinks and data points with abnormally high speed (over 40 pixels per ms, approximately 650 degrees per second) the $x$ and $y$ distributions of the Waldo task data were used to automatically correct miscalibrations. Using 100 bins in each direction, (allowing 15\% on either side of the nominal screen boundaries) the final bin with a cumulative frequency under $0.5\%$ and the first bin with a cumulative frequency over $99.5\%$ were used as the data edges, and aligned to the screen dimensions by rescaling. Data points outside this region were then excluded. As a final cleaning stage, short segments below 20ms in length) were also discarded.

Participants with more than 20\% of the 1000Hz data missing after preprocessing in either task were excluded (4 out of 35). The remaining participants had a mean of 1.1\% missing fixations in the Waldo task and a mean of 1.0\% missing fixations in the random pixel task (maximum 5.7\% and 8.7\% respectively). For remaining fixations, the landing times and the means of valid $x$ and $y$ positions were extracted.

\section{Time constant optimization}
\begin{figure}
\centering
\includegraphics[width=0.45\textwidth]{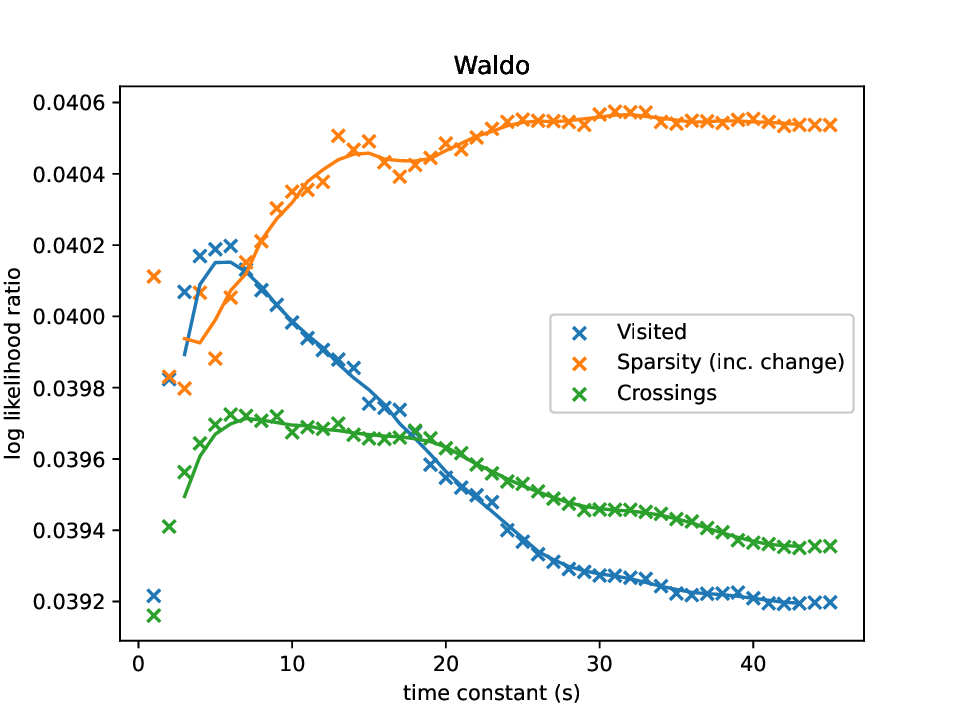}
\includegraphics[width=0.44\textwidth,trim=12 0 0 0,clip]{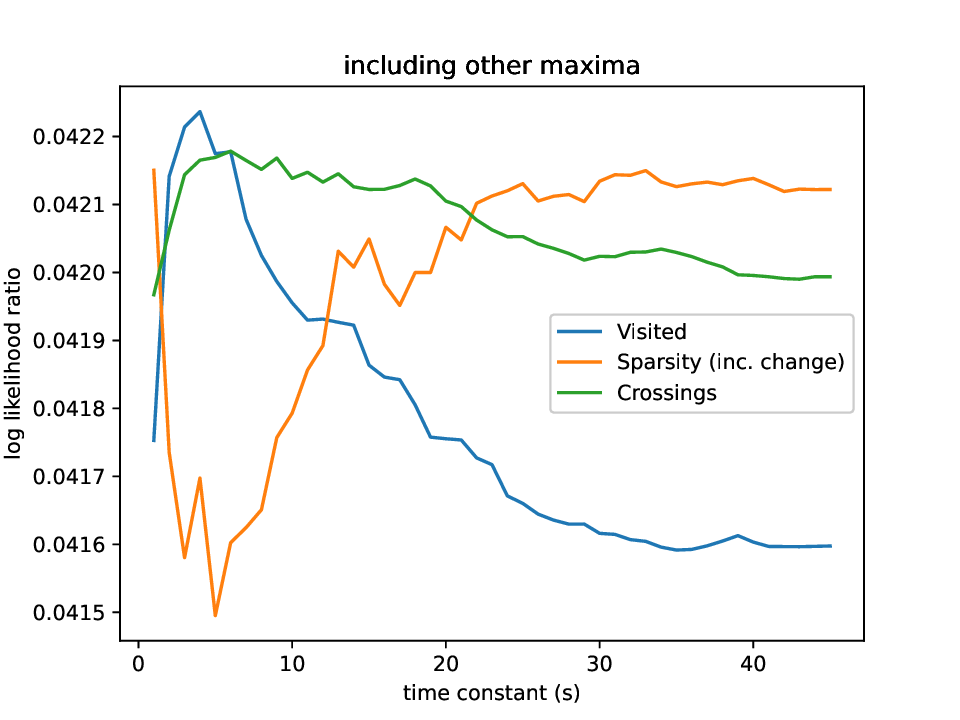}\\
\includegraphics[width=0.45\textwidth]{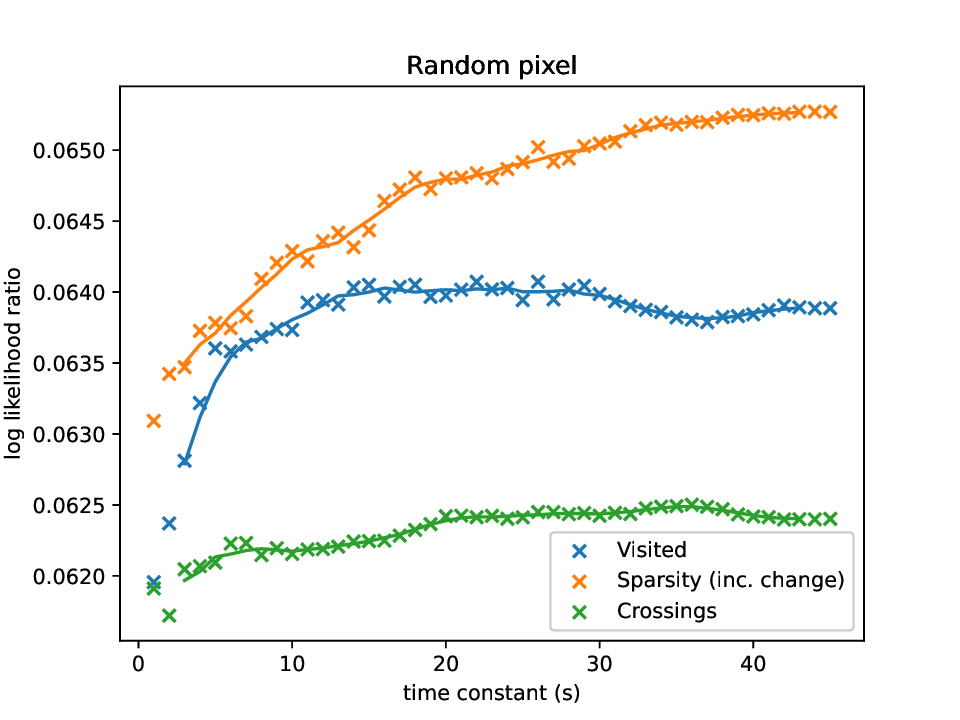}
\includegraphics[width=0.44\textwidth,trim=10 0 0 0,clip]{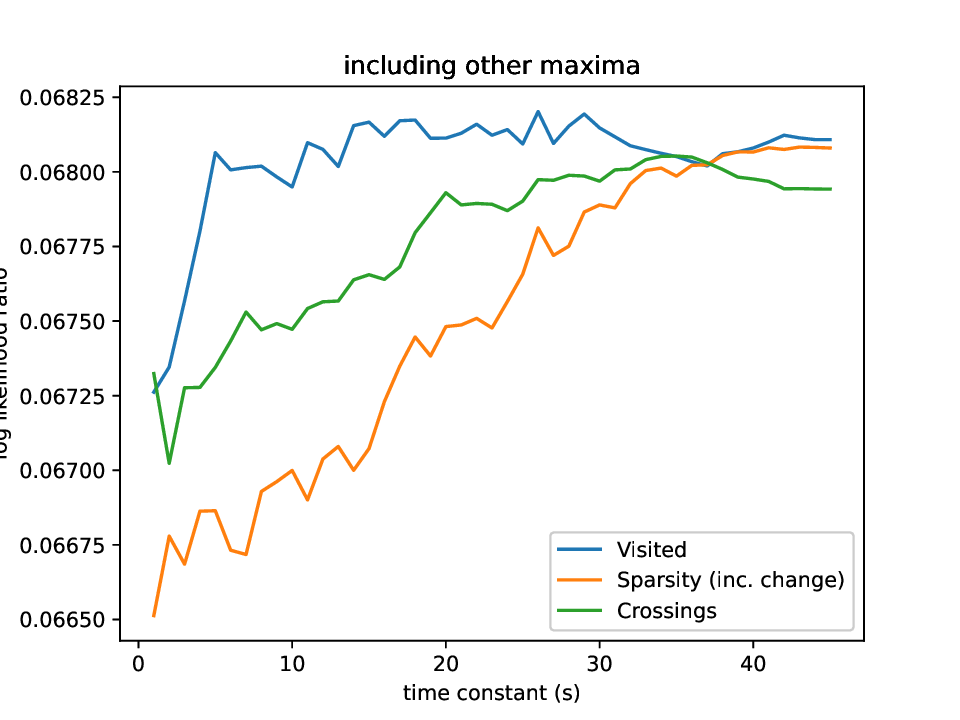}
\caption{Time constant optimization for the Waldo task (above) and the random pixel task (below). Left panels: Log-likelihood ratio vs $\tau$ for each of the three variable-history predictors when only one is included. The curve applies an average over a moving window of five seconds width. Right panels: The same time parameter sweeps, but including the maximum of the other two predictors. \label{fig:timeopt}}
\end{figure}

Figure \ref{fig:timeopt} shows the time constant optimization curves for both tasks. The peaks of the smoothed curves for the individual predictors were then used as the final time constants. This was done to preclude the possibility of further unmodeled history dependence that could affect a joint maximum.

When including the optima for the two other predictors, the curves peak at approximately the same height, and with only minor changes in shape.

\subsection{Joint optimization} 
To further check for unseen dependencies between the chosen time-variable predictors, the DIRECT algorithm was used to determine the global optimum \cite{jones1993lipschitzian} as implemented in {\em SciPy 1.11.4}. A minimum proportional volume of $10^{-5}$ was used as a stopping criterion.

\section{Nonparametric model}
This model is based on sampling from the empirical distribution
\begin{equation}
P(\Delta x_t,\Delta y_t|x_t,y_t)
\end{equation}
where this is represented as a four dimensional histogram, normalized so that it sums to one for all pairs $x_t,y_t$. 50 bins were used in the $x,y$ dimensions, and 100 in the $\Delta_x$ and $\Delta_y$ dimensions. Prior to normalization, gaussian kernel smoothing was applied. Two filter widths were used, $\sigma_{X,Y}$ for the $x_t,y_t$ dimensions and $\sigma_{dX,dY}$ for the $\Delta_x$ and $\Delta_y$ dimensions, which were jointly optimized as detailed below.

\subsection{Kernel smoothing optimization}

For each combination of the smoothing parameters, a conditional distribution was obtained from the training data. A step selection process was then run on the test data, generating control points using the unconditional distribution $P(\Delta x_t,\Delta y_t)$ from the training data. The conditional probability in logits was then calculated from each smoothed conditional distribution for each case and control point. The log likelihood was then calculated by fitting the logistic regression model for each combination of $\sigma_{X,Y}$ and $\sigma_{dX,dY}$ in turn.

\begin{figure}
\centering
\includegraphics[width=0.7\textwidth]{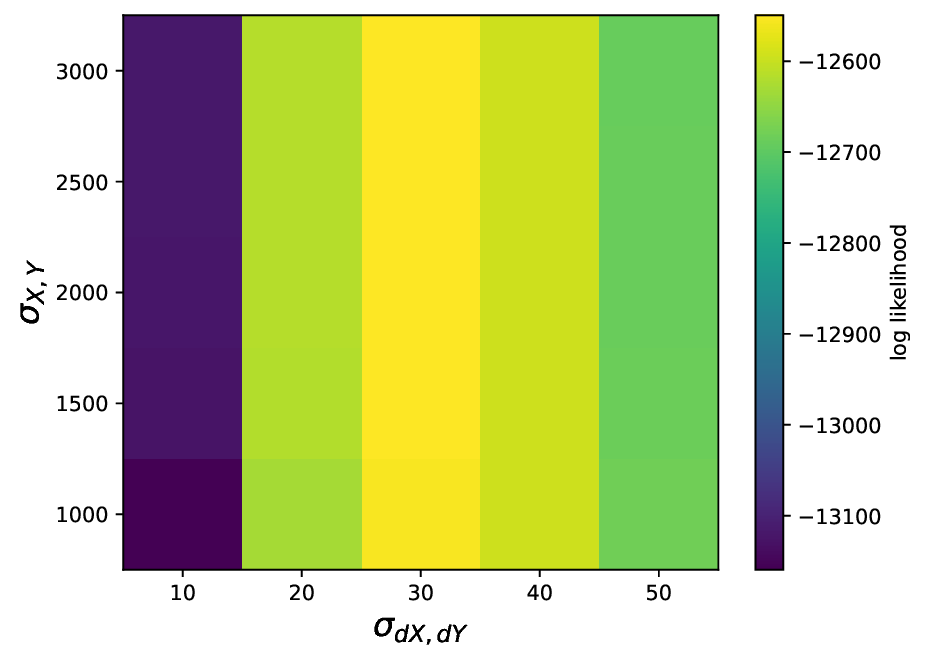}
\caption{Optimization of the KDE smoothing in the nonparametric model. \label{fig:sigma}}
\end{figure}

As shown in Figure \ref{fig:sigma}, the optimal filter size is comparable to the screen dimensions, making the final distribution close to the unconditional one.

\subsection{Amplitude spline}
\begin{figure}
\centering
\includegraphics[width=0.8\textwidth]{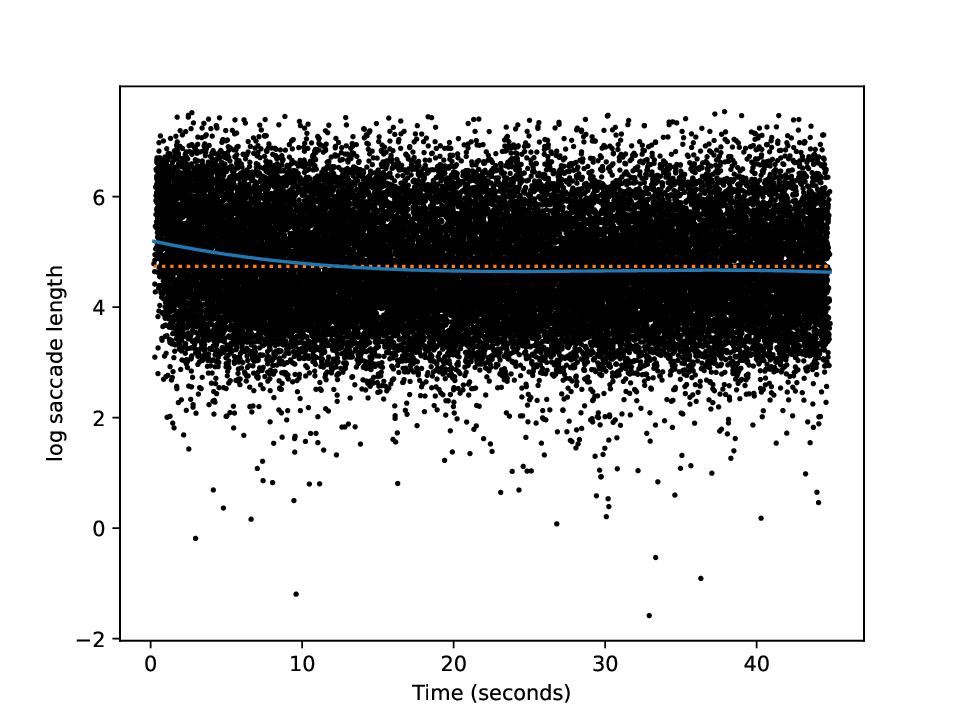}\\
\includegraphics[width=0.7\textwidth]{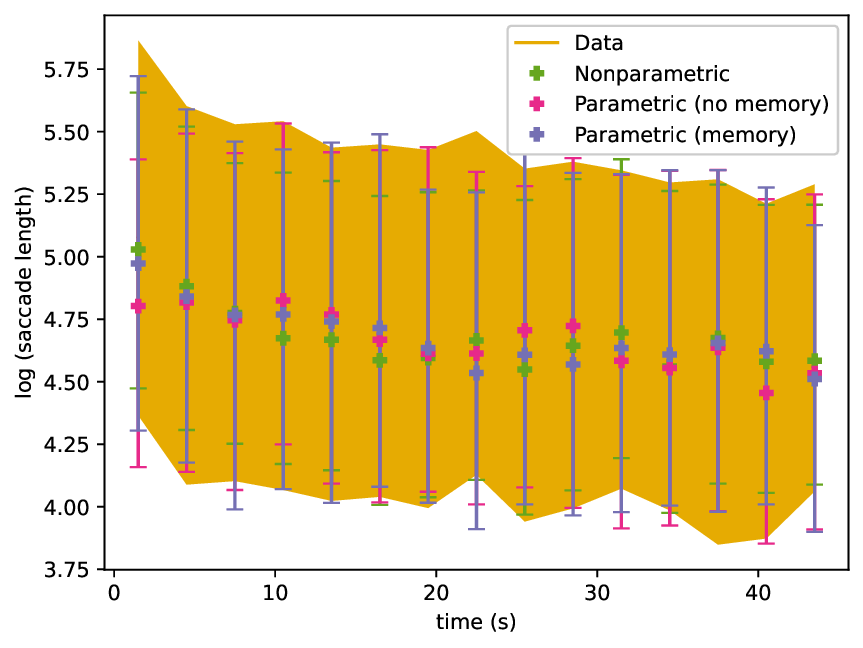}
\caption{Above: Fitting of the smoothing spline for the non-parametric model. Points are all saccade log-lengths in pixels from the training data. The solid line is the smoothing spline, and the dotted line is the overall mean. Below:  Time evolution of mean log distance vs time for the data and the final models. Plotted are the median (line and points) and the interquartile range (region and errorbars). \label{fig:ampspline}}
\end{figure}

For the final fitting of the nonparametric model, the $x$ and $y$ displacements of sampled saccades were multiplied by a scaling factor $\mathrm{exp}(a_0 - a(t))$, where $a(t)$ was an estimate of the mean log saccade length over time $t$ during a trial and $a_0$ was the overall mean log-length. The function $a(t)$ was obtained with a cubic spline fitted to an aggregated scatter plot of log-length against time in the trial, for all trials in the training data (see Figure \ref{fig:ampspline}). The {\em SciPy} function {\em splrev} was used to obtain the spline, with smoothing factor equal to the variance of log-length multiplied by the number of points. The mean and interquartile ranges of saccade log-lengths from the simulations, corresponding to Figure 3 in the main text, are also shown in Figure \ref{fig:ampspline}, along with the data and parametric models for comparison.

For the generative model, increments were first sampled based on the conditional distribution, then multiplied by the inverse of the scaling factor, i.e.\ $\mathrm{exp}(a(t)-a_0)$. Fixations outside the screen region were discarded and resampled.

\section{Parametric model}

\subsection{Variable importance}

The full parametric (logistic regression) models included the predictors in Figure 1 of the main text, as well as the linear saccade length, square of the log saccade length, initial distance from the center, all second-order interactions of these terms, plus the interaction between log-length and time to capture the coarse-to-fine transition.

\begin{figure}
\centering
\includegraphics[width=0.45\textwidth]{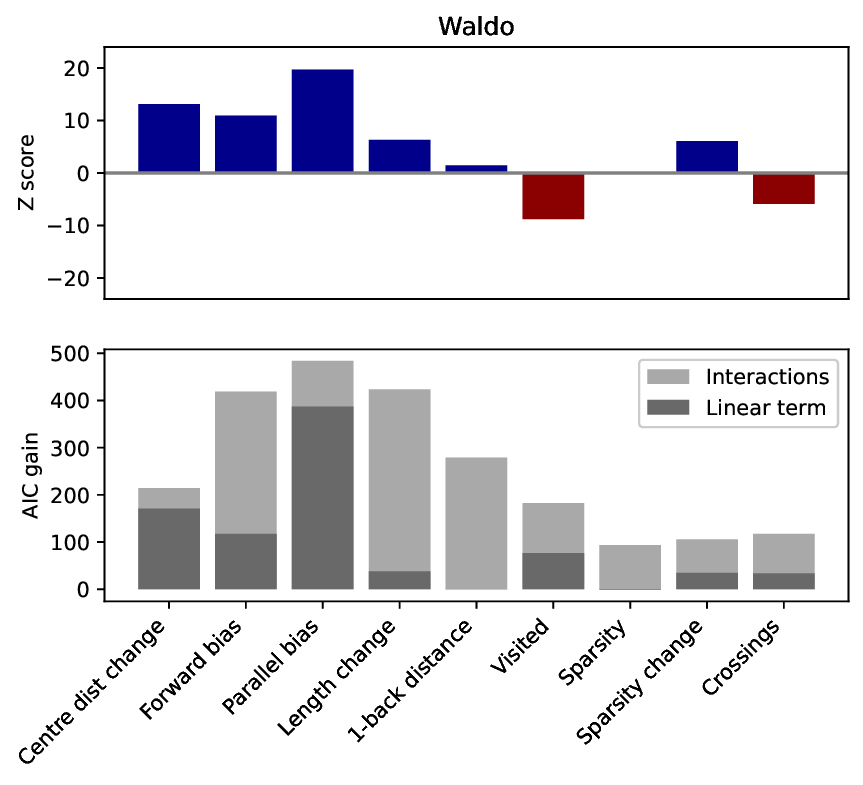}
\includegraphics[width=0.45\textwidth]{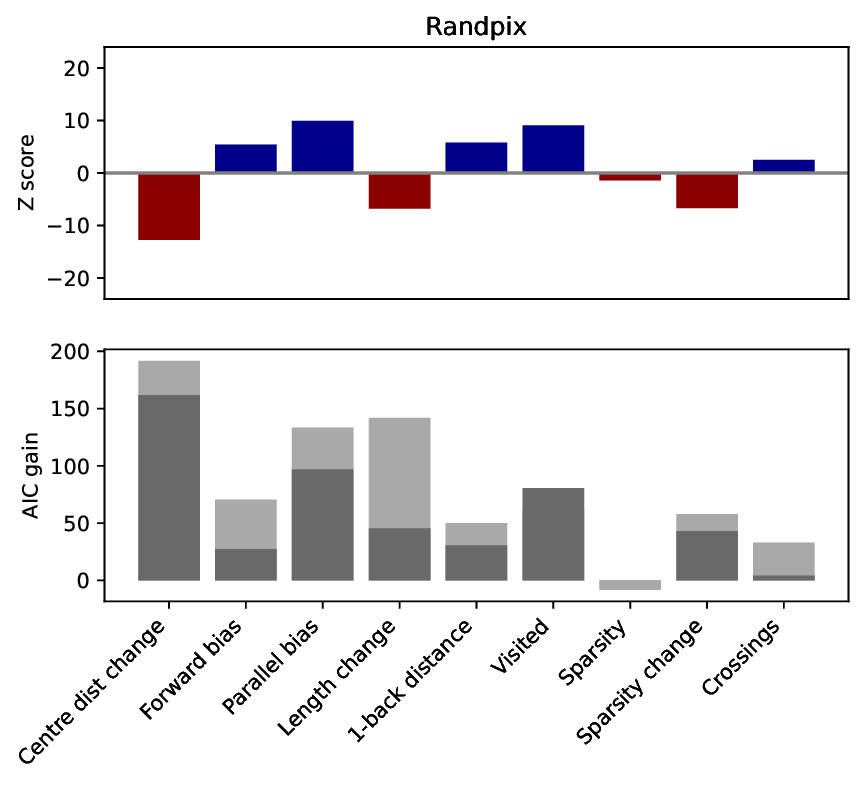}
\caption{Normalized effect sizes ($z$ score) and AIC gains for each predictor. Z-scores were set to zero if the AIC gain for the linear model was negative.\label{fig:aic}}
\end{figure}

To assess variable importance and linear effect sizes in the parametric model, we fit two models for each predictor: one with the predictor excluded, and one with only this predictor's interactions excluded. The AIC values obtained were each compared to that for the full model to determine the relative importance of linear terms and interactions. Z-scores were obtained for each model with the predictor's interactions removed. Figure \ref{fig:aic} shows these z-scores and relative AIC contributions of each predictor, for both tasks.

\subsection{Spatial dependence}

To quantify the dependence on screen position, we used the change in sparsity ($\Delta s' = s' - s_0$) and change in distance from the centre ($\Delta c' = c' - c_0$) with a new candidate fixation, as well as the final sparsity $s'$ and initial distance from the centre $c_0$. A logistic regression model was fit using these four terms and their interactions.

The gradients were then computed from the coefficients $\beta$ as:
\begin{eqnarray}
\frac{\partial L(c',s')}{\partial c'}\bigg\rvert_{c'=c_0,s'=s_0} &=& \beta(c_0) + c\beta(c_0 : \Delta c) + s\beta(c_0:\Delta s) \\
\frac{\partial L(c',s')}{\partial s'}\bigg\rvert_{c'=c_0,s'=s_0} &=& \beta(\Delta s') + \beta(s') + c[\beta(c_0:s')+\beta(c_0:\Delta s)] + s\beta(\Delta s:s')
\end{eqnarray}
where $L(c,s)$ is the logit function and $:$ denotes interaction terms.

\subsection{Self-crossings}

\begin{figure}
\centering
\includegraphics[width=0.45\textwidth]{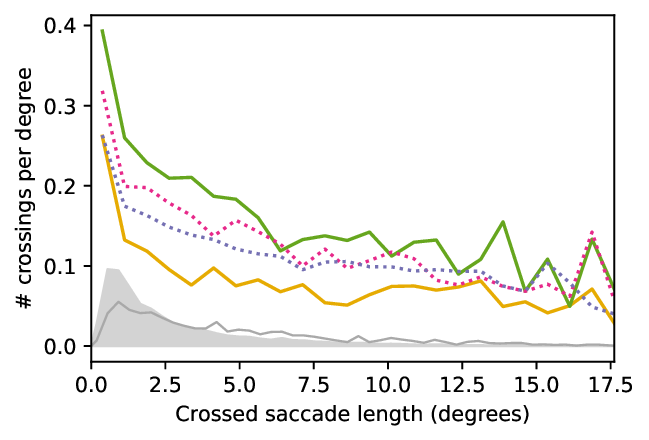}
\includegraphics[width=0.45\textwidth]{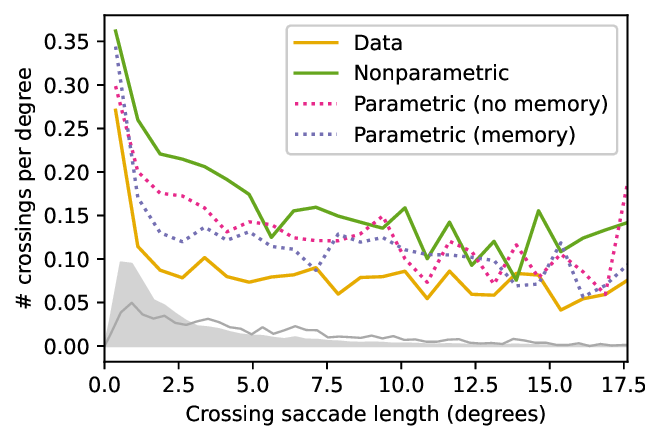}
\caption{Expected number of crossings per pixel for (left) the first saccade and (right) second saccade in a crossing pair, averaged over all saccades. \label{fig:perpix}}
\end{figure}

Figure 3 of the main text contains saccade statistics that include both saccades for every self-crossing. When including only the first or only the second saccade in a crossing pair, the statistics are largely similar, as can be seen in Figure \ref{fig:perpix}.

\subsection{Stepwise regression}

For the final parametric models (main text Figure 1 and Figure 3), all second-order interactions as well as $\mathrm{log}(r)*t$ were initially included, and stepwise backward elimination was used to exclude terms beginning with the highest p-value. A term's interaction with the log-length was not excluded if there was an existing interaction with the square of the log-length. The process is stopped at the first increase of the AIC. For the Waldo task, this reduced the final model from 120 to 86 terms. Using the same procedure on the random pixel task yielded 65 terms in the final model.

Outputs of the final model fits can be found at the end of this document in datasets \ref{data:waldo} and \ref{data:randpix}. 

\section{Mixed-effect modelling}
To limit the model's degrees of freedom, the mixed-effects logistic regression contained only linear fixed effects corresponding to the 12 predictors in Figure 1 of the main text, as well as the square of the log saccade distance. The training data was normalized to a mean of zero and a standard deviation of one for each predictors.

\begin{figure}
\centering
\includegraphics[width=0.45\textwidth]{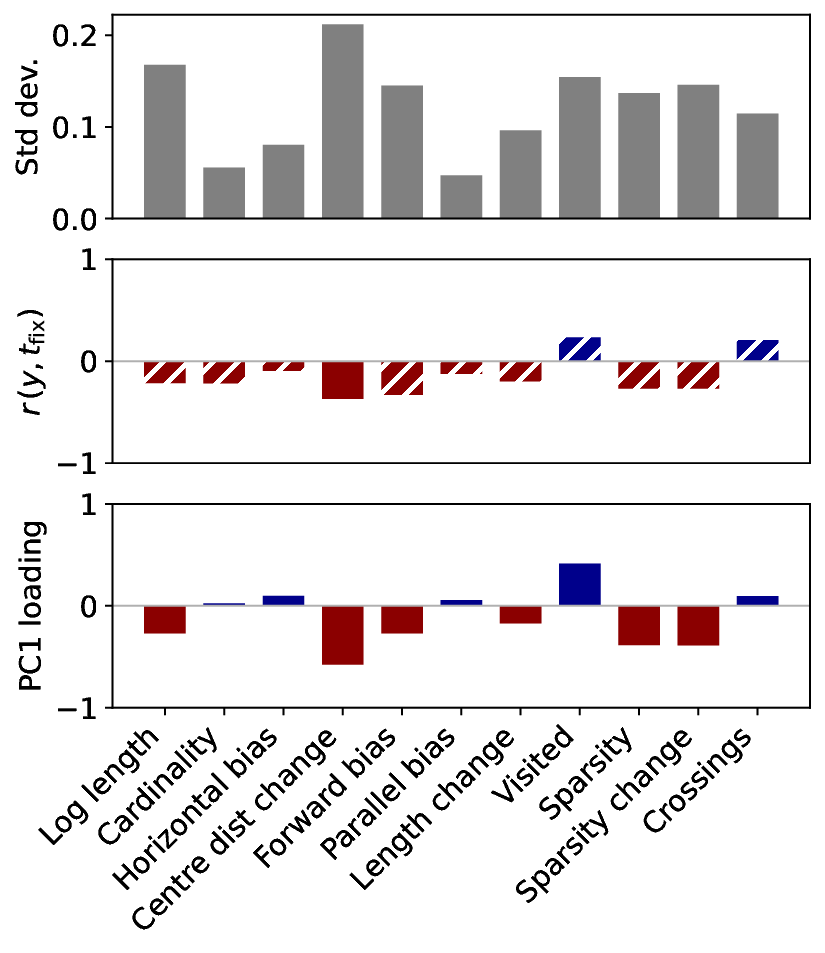}
\caption{Individual differences from the random pixel task. Top: Fitted standard deviation of random effects. Middle: Pearson product-moment correlations with means of the trial-level medians of fixation duration. Bottom: Loadings of the first principal component.\label{fig:randpix_ind}}
\end{figure}

Figure \ref{fig:randpix_ind} shows the significant individual differences for the random pixel task, equivalent to Figure 4 of the main text. Since the trial-level random effect was not fit, the intraclass correlation coefficient (ICC) was not estimated, and more significant predictors were also found for this reason.

\begin{figure}
\centering
\includegraphics[width=0.45\textwidth]{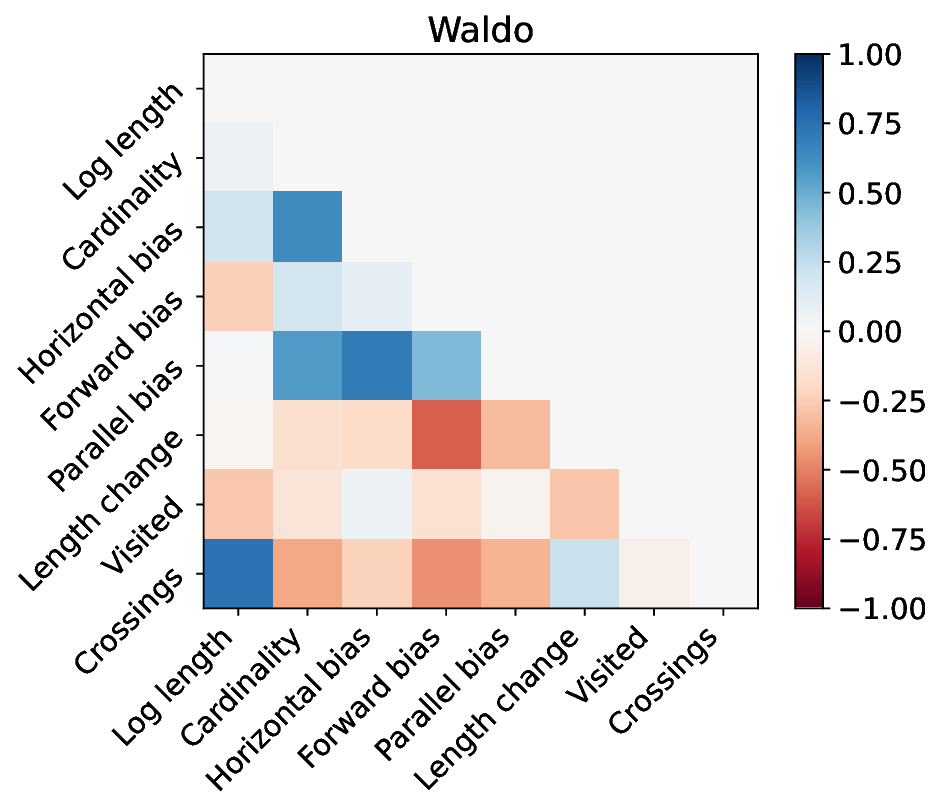}
\includegraphics[width=0.45\textwidth]{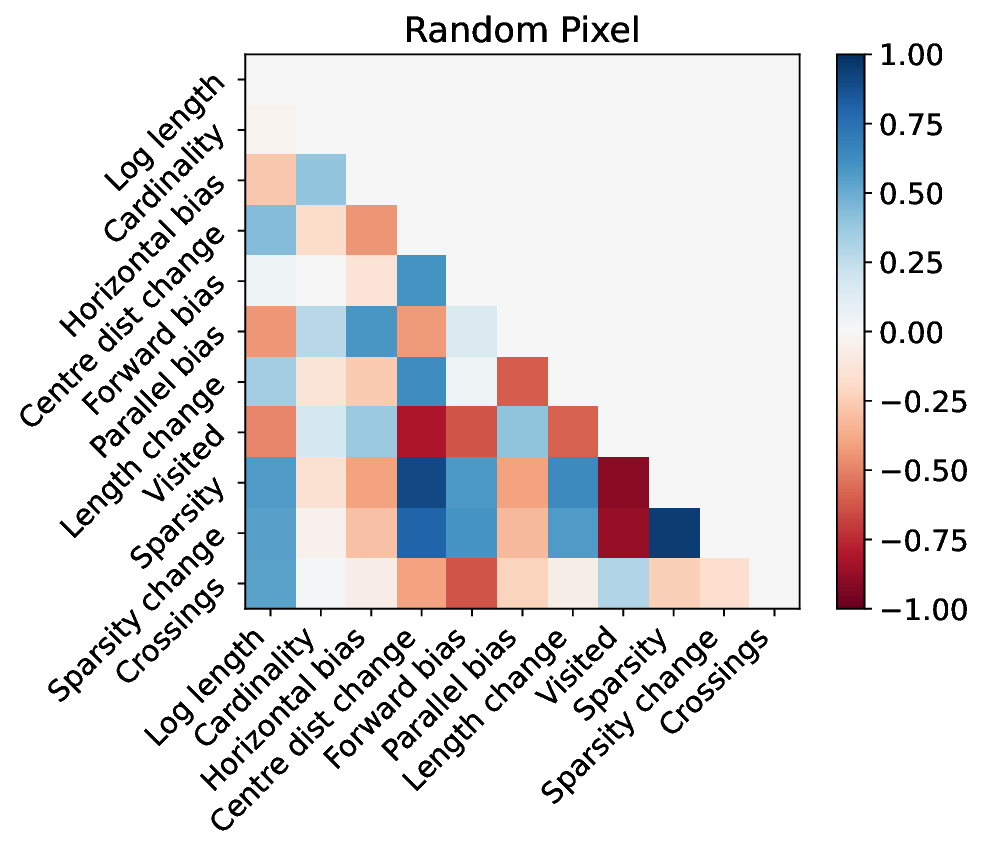}
\caption{Correlation matrices for fitted random effects for the significant predictors in each task. Note that the participant effect was not tested against a trial effect only model for the random pixel task. \label{fig:corrmat}}
\end{figure}

Figure \ref{fig:corrmat} shows the matrix of Pearson product-moment correlations between all pairs of random effects in both tasks.

\begin{figure}
\centering
\includegraphics[width=0.8\textwidth]{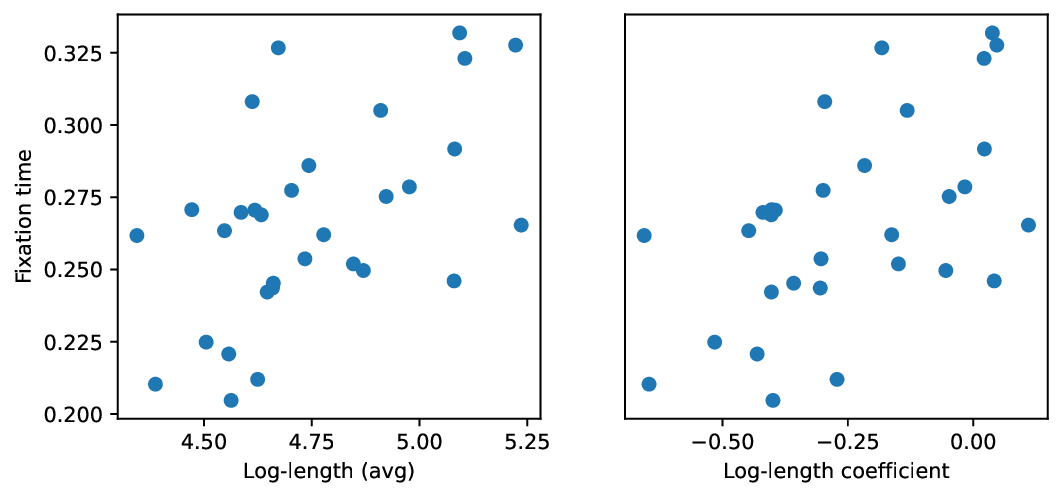}
\caption{Participant-level measures of fixation duration plotted against two participant-level measures of log saccade length for the Waldo task. The panel on the left uses a simple median of all saccade log-lengths per trial, with a mean then taken over the 9 trials. The panel on the right uses the random effect coefficients obtained for the {\em log length} predictor in the mixed-effect logistic regression. These are not centered at zero due to the addition of the fixed effect of {\em log length}. \label{fig:fixtime}}
\end{figure}

For the Waldo task, the mean of the trial-median fixation duration was well correlated with the log saccade length, whether the latter was measured by the mean of the trial-median ($r=0.53, p=0.0020$) or the random coefficient obtained from the logistic regression ($r=0.55, p=0.0015$). These correlations can be seen in Figure \ref{fig:fixtime}.

\subsection{Principal component analysis}
\begin{figure}
\centering
\includegraphics[width=0.6\textwidth]{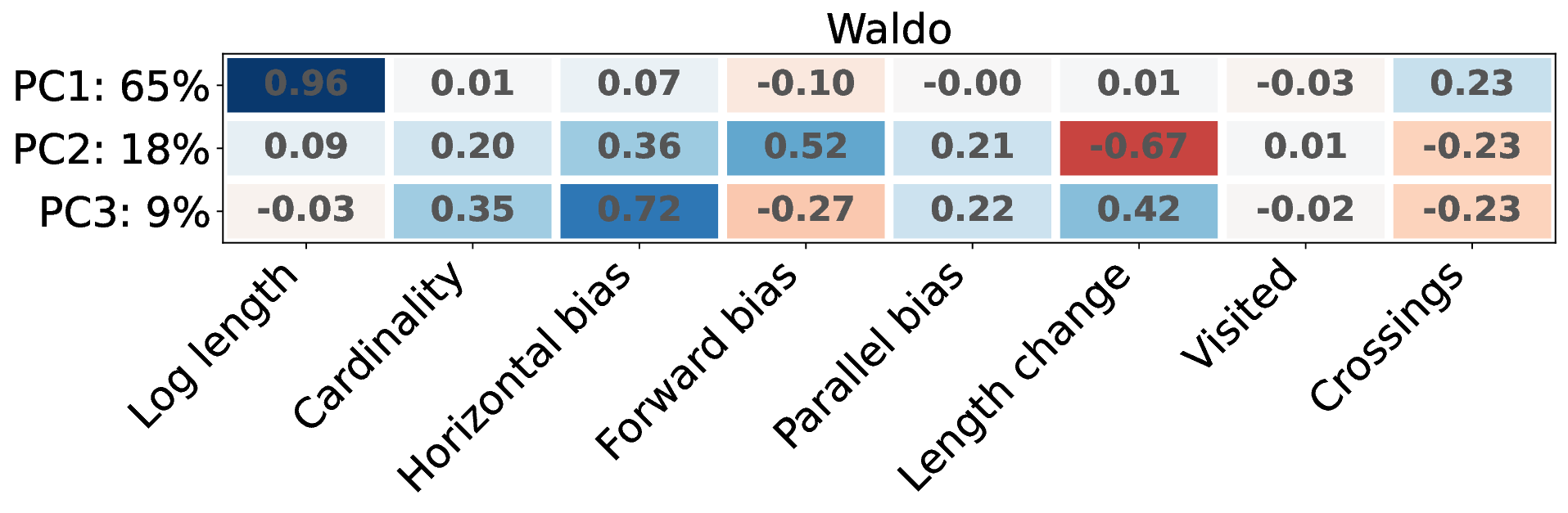}\\
\includegraphics[width=0.6\textwidth]{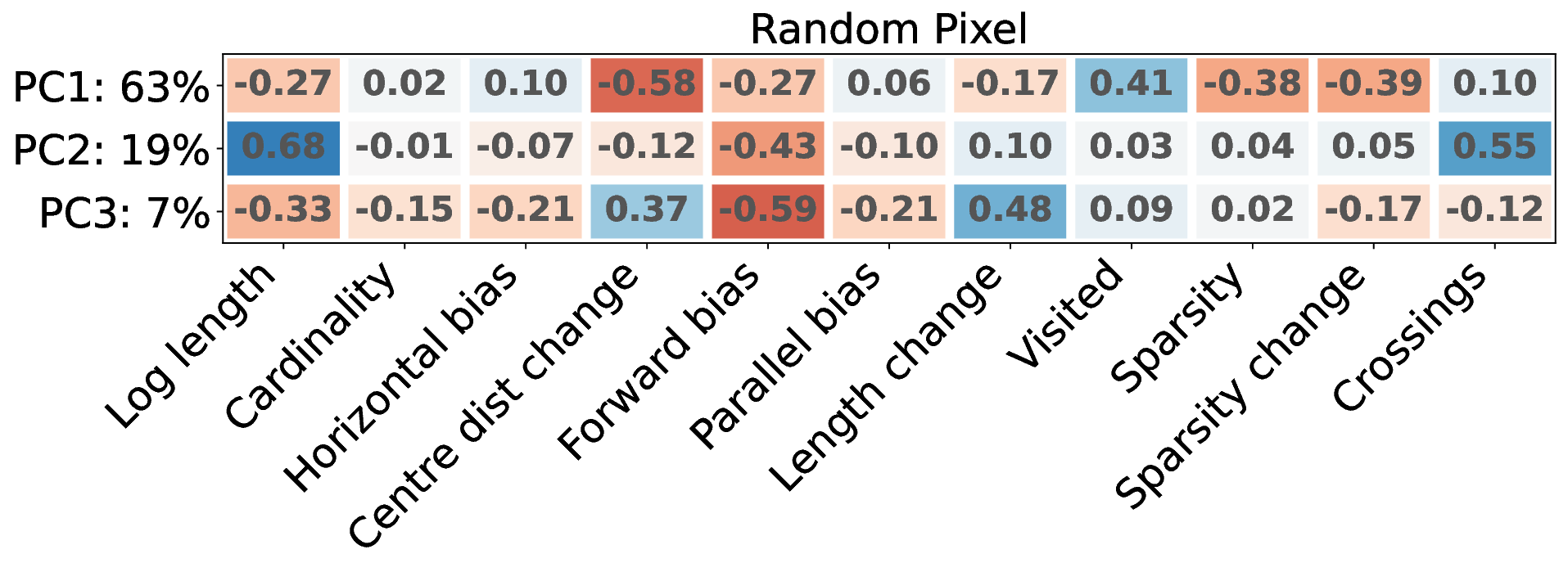}
\caption{Loadings of the first three principal components for each task. The variance explained is shown on the left of each row. \label{fig:pc_loadings}}
\end{figure}

PCA was run using the {\em PCA} class in {\em scikit-learn 1.2.2}. Since the variables were standardized, the resulting variances can be used as an estimate of each variable's importance. These were therefore not normalized again prior to principal component analysis. Figure \ref{fig:pc_loadings} shows the loadings of the first three principal components for each task.




\FloatBarrier

\section*{Dataset 1: Final logistic regression model for the Waldo task. \label{data:waldo}}

Glossary: \texttt{dir\_change} is forward bias, \texttt{dir\_change2} is parallel bias, and \texttt{whistdist} is sparsity.

{\footnotesize
\begin{verbatim}
                           Logit Regression Results                           
==============================================================================
Dep. Variable:                 target   No. Observations:               280230
Model:                          Logit   Df Residuals:                   280141
Method:                           MLE   Df Model:                           88
Date:                Wed, 15 Oct 2025   Pseudo R-squ.:                 0.06531
Time:                        13:39:51   Log-Likelihood:                -85149.
converged:                       True   LL-Null:                       -91098.
Covariance Type:            nonrobust   LLR p-value:                     0.000
=====================================================================================================
                                        coef    std err          z      P>|z|      [0.025      0.975]
-----------------------------------------------------------------------------------------------------
Intercept                             3.1740      1.636      1.940      0.052      -0.033       6.381
dist                                 -0.6364      0.060    -10.682      0.000      -0.753      -0.520
logdist                              -2.6583      0.859     -3.095      0.002      -4.342      -0.975
logdist_sq                            0.9270      0.154      6.005      0.000       0.624       1.230
cardinality                           0.3199      0.187      1.707      0.088      -0.047       0.687
horiz_align                          -0.1839      0.121     -1.518      0.129      -0.421       0.054
centre_dist_change                    0.0126      0.005      2.425      0.015       0.002       0.023
dir_change                           -0.0892      0.193     -0.461      0.645      -0.468       0.290
dir_change2                           0.4489      0.068      6.597      0.000       0.316       0.582
log_amp_change                       -1.0185      0.123     -8.302      0.000      -1.259      -0.778
dist_1back                           -0.0003      0.002     -0.185      0.853      -0.004       0.003
centre_dist_init                     -0.0013      0.001     -1.725      0.084      -0.003       0.000
have_visited                         -0.2071      0.456     -0.454      0.650      -1.101       0.687
whistdist                             2.5897      2.504      1.034      0.301      -2.317       7.497
whistdist_change                    -19.0794      7.268     -2.625      0.009     -33.324      -4.835
crossings                             2.3741      0.279      8.519      0.000       1.828       2.920
dist:logdist                          0.1537      0.014     10.676      0.000       0.126       0.182
dist:logdist_sq                      -0.0096      0.001    -10.523      0.000      -0.011      -0.008
dist:horiz_align                      0.0003      0.000      1.928      0.054    -4.5e-06       0.001
dist:centre_dist_change           -4.349e-06   1.51e-06     -2.877      0.004   -7.31e-06   -1.39e-06
dist:dir_change2                     -0.0004   8.13e-05     -5.258      0.000      -0.001      -0.000
dist:log_amp_change                  -0.0004      0.000     -3.800      0.000      -0.001      -0.000
dist:dist_1back                    3.696e-06   8.76e-07      4.220      0.000    1.98e-06    5.41e-06
dist:centre_dist_init              4.251e-06   9.81e-07      4.332      0.000    2.33e-06    6.17e-06
dist:have_visited                     0.0007      0.000      1.598      0.110      -0.000       0.002
dist:whistdist                       -0.0073      0.002     -3.076      0.002      -0.012      -0.003
logdist:cardinality                  -0.3256      0.081     -4.043      0.000      -0.483      -0.168
logdist:horiz_align                   0.1764      0.022      8.153      0.000       0.134       0.219
logdist:centre_dist_change           -0.0054      0.002     -2.523      0.012      -0.010      -0.001
logdist:dir_change                    0.2912      0.082      3.534      0.000       0.130       0.453
logdist:log_amp_change                0.2252      0.024      9.337      0.000       0.178       0.272
logdist:dist_1back                    0.0012      0.001      1.771      0.077      -0.000       0.003
logdist:centre_dist_init              0.0007      0.000      1.734      0.083   -9.08e-05       0.001
logdist:have_visited                  0.2551      0.234      1.092      0.275      -0.203       0.713
logdist:whistdist                    -2.7760      1.267     -2.191      0.028      -5.259      -0.293
logdist:whistdist_change              7.4551      2.657      2.805      0.005       2.247      12.664
logdist:crossings                    -1.0166      0.111     -9.176      0.000      -1.234      -0.799
logdist_sq:cardinality                0.0617      0.009      6.901      0.000       0.044       0.079
logdist_sq:centre_dist_change         0.0006      0.000      2.401      0.016       0.000       0.001
logdist_sq:dir_change                -0.0435      0.009     -4.793      0.000      -0.061      -0.026
logdist_sq:dist_1back                -0.0002    8.4e-05     -2.885      0.004      -0.000   -7.77e-05
logdist_sq:centre_dist_init          -0.0001   5.63e-05     -1.828      0.068      -0.000    7.45e-06
logdist_sq:have_visited              -0.0600      0.031     -1.914      0.056      -0.121       0.001
logdist_sq:whistdist                  0.5178      0.169      3.069      0.002       0.187       0.849
logdist_sq:whistdist_change          -0.6005      0.243     -2.469      0.014      -1.077      -0.124
logdist_sq:crossings                  0.1053      0.011      9.167      0.000       0.083       0.128
cardinality:horiz_align               0.1592      0.020      8.140      0.000       0.121       0.198
cardinality:dir_change                0.0748      0.014      5.166      0.000       0.046       0.103
cardinality:dir_change2               0.1001      0.014      6.961      0.000       0.072       0.128
cardinality:log_amp_change            0.0348      0.015      2.383      0.017       0.006       0.063
cardinality:dist_1back               -0.0002   5.99e-05     -3.918      0.000      -0.000      -0.000
cardinality:centre_dist_init          0.0002   3.95e-05      4.412      0.000    9.69e-05       0.000
cardinality:whistdist_change         -0.7543      0.203     -3.713      0.000      -1.152      -0.356
cardinality:crossings                -0.0304      0.016     -1.880      0.060      -0.062       0.001
horiz_align:centre_dist_change        0.0002   8.66e-05      2.035      0.042    6.52e-06       0.000
horiz_align:dir_change               -0.0655      0.015     -4.382      0.000      -0.095      -0.036
horiz_align:dir_change2              -0.0197      0.013     -1.466      0.143      -0.046       0.007
horiz_align:log_amp_change            0.0507      0.014      3.508      0.000       0.022       0.079
horiz_align:dist_1back               -0.0002   7.31e-05     -3.073      0.002      -0.000   -8.14e-05
horiz_align:centre_dist_init         -0.0004   3.92e-05     -9.062      0.000      -0.000      -0.000
horiz_align:have_visited              0.0651      0.024      2.691      0.007       0.018       0.113
horiz_align:whistdist                -0.3802      0.140     -2.712      0.007      -0.655      -0.105
horiz_align:whistdist_change          0.4121      0.227      1.814      0.070      -0.033       0.857
horiz_align:crossings                -0.0407      0.017     -2.466      0.014      -0.073      -0.008
centre_dist_change:dir_change        -0.0003   7.43e-05     -3.573      0.000      -0.000      -0.000
centre_dist_change:log_amp_change     0.0002   5.51e-05      3.038      0.002    5.94e-05       0.000
centre_dist_change:dist_1back      8.676e-07   3.22e-07      2.698      0.007    2.37e-07     1.5e-06
centre_dist_change:have_visited      -0.0002      0.000     -1.641      0.101      -0.000    4.22e-05
centre_dist_change:whistdist          0.0031      0.001      5.501      0.000       0.002       0.004
dir_change:log_amp_change            -0.2162      0.016    -13.365      0.000      -0.248      -0.185
dir_change:dist_1back                 0.0007   6.62e-05     10.751      0.000       0.001       0.001
dir_change:have_visited              -0.1211      0.023     -5.169      0.000      -0.167      -0.075
dir_change2:log_amp_change           -0.0853      0.014     -6.307      0.000      -0.112      -0.059
dir_change2:dist_1back                0.0004   7.09e-05      5.913      0.000       0.000       0.001
dir_change2:whistdist                -0.3917      0.129     -3.043      0.002      -0.644      -0.139
dir_change2:crossings                 0.0468      0.017      2.834      0.005       0.014       0.079
log_amp_change:dist_1back            -0.0004   7.48e-05     -5.654      0.000      -0.001      -0.000
log_amp_change:centre_dist_init    8.637e-05   3.99e-05      2.165      0.030    8.18e-06       0.000
log_amp_change:whistdist              0.1903      0.114      1.667      0.096      -0.033       0.414
dist_1back:have_visited               0.0008      0.000      7.226      0.000       0.001       0.001
dist_1back:crossings                 -0.0002   6.89e-05     -2.864      0.004      -0.000   -6.23e-05
centre_dist_init:have_visited        -0.0002   6.73e-05     -3.561      0.000      -0.000      -0.000
centre_dist_init:whistdist_change     0.0014      0.001      2.361      0.018       0.000       0.003
centre_dist_init:crossings           -0.0001   4.69e-05     -2.635      0.008      -0.000   -3.16e-05
have_visited:whistdist_change         2.4978      0.442      5.646      0.000       1.631       3.365
have_visited:crossings                0.0586      0.027      2.151      0.031       0.005       0.112
whistdist:whistdist_change           -5.4010      0.909     -5.940      0.000      -7.183      -3.619
t                                     0.0251      0.003      9.476      0.000       0.020       0.030
logdist:t                            -0.0049      0.001     -8.646      0.000      -0.006      -0.004
=====================================================================================================
\end{verbatim}
}
\newpage
\section*{Dataset 2: Final logistic regression model for the random pixel task \label{data:randpix}}
{\footnotesize
\begin{verbatim}
                           Logit Regression Results                           
==============================================================================
Dep. Variable:                 target   No. Observations:                59990
Model:                          Logit   Df Residuals:                    59922
Method:                           MLE   Df Model:                           67
Date:                Wed, 15 Oct 2025   Pseudo R-squ.:                  0.1047
Time:                        13:56:21   Log-Likelihood:                -17460.
converged:                       True   LL-Null:                       -19502.
Covariance Type:            nonrobust   LLR p-value:                     0.000
=======================================================================================================
                                          coef    std err          z      P>|z|      [0.025      0.975]
-------------------------------------------------------------------------------------------------------
Intercept                               3.3689      2.052      1.642      0.101      -0.653       7.391
dist                                    0.0120      0.006      2.168      0.030       0.001       0.023
logdist                                -2.1800      0.934     -2.335      0.020      -4.010      -0.350
logdist_sq                              0.1783      0.120      1.485      0.137      -0.057       0.414
cardinality                             0.1995      0.098      2.029      0.042       0.007       0.392
horiz_align                            -1.7282      0.808     -2.140      0.032      -3.311      -0.145
centre_dist_change                      0.0540      0.013      4.039      0.000       0.028       0.080
dir_change                             -0.1900      0.220     -0.863      0.388      -0.622       0.242
dir_change2                            -0.6873      0.525     -1.310      0.190      -1.716       0.341
log_amp_change                         -1.9263      0.206     -9.346      0.000      -2.330      -1.522
dist_1back                              0.0039      0.001      3.042      0.002       0.001       0.006
centre_dist_init                        0.0026      0.001      1.773      0.076      -0.000       0.005
have_visited                            0.2802      0.041      6.757      0.000       0.199       0.362
whistdist                               8.5575      4.011      2.133      0.033       0.695      16.420
whistdist_change                        3.9141      2.596      1.508      0.132      -1.174       9.002
crossings                               0.4921      0.224      2.197      0.028       0.053       0.931
dist:logdist_sq                        -0.0002   8.75e-05     -2.735      0.006      -0.000   -6.78e-05
dist:cardinality                        0.0009      0.000      7.195      0.000       0.001       0.001
dist:horiz_align                        0.0017      0.001      2.402      0.016       0.000       0.003
dist:centre_dist_change             -1.496e-05   3.29e-06     -4.541      0.000   -2.14e-05    -8.5e-06
dist:dir_change                        -0.0005      0.000     -1.641      0.101      -0.001    8.91e-05
logdist:horiz_align                     0.9550      0.403      2.369      0.018       0.165       1.745
logdist:centre_dist_change             -0.0236      0.005     -4.344      0.000      -0.034      -0.013
logdist:dir_change                      0.0940      0.051      1.834      0.067      -0.006       0.194
logdist:dir_change2                     0.5527      0.215      2.571      0.010       0.131       0.974
logdist:log_amp_change                  0.3494      0.043      8.139      0.000       0.265       0.434
logdist:dist_1back                     -0.0007      0.000     -3.064      0.002      -0.001      -0.000
logdist:centre_dist_init               -0.0016      0.001     -2.529      0.011      -0.003      -0.000
logdist:whistdist                      -3.7504      1.685     -2.226      0.026      -7.053      -0.448
logdist:whistdist_change               -1.4655      0.495     -2.962      0.003      -2.435      -0.496
logdist:crossings                      -0.1472      0.083     -1.772      0.076      -0.310       0.016
logdist_sq:horiz_align                 -0.1087      0.053     -2.070      0.038      -0.212      -0.006
logdist_sq:centre_dist_change           0.0026      0.001      4.505      0.000       0.001       0.004
logdist_sq:dir_change2                 -0.0692      0.023     -3.065      0.002      -0.114      -0.025
logdist_sq:centre_dist_init             0.0002   6.91e-05      2.738      0.006    5.38e-05       0.000
logdist_sq:whistdist                    0.3935      0.175      2.245      0.025       0.050       0.737
logdist_sq:crossings                    0.0134      0.008      1.687      0.092      -0.002       0.029
cardinality:horiz_align                 0.1122      0.043      2.622      0.009       0.028       0.196
cardinality:whistdist                  -0.3410      0.206     -1.658      0.097      -0.744       0.062
horiz_align:centre_dist_change          0.0003      0.000      1.941      0.052   -3.27e-06       0.001
horiz_align:dir_change                 -0.1263      0.037     -3.438      0.001      -0.198      -0.054
horiz_align:dist_1back                 -0.0003      0.000     -2.124      0.034      -0.001   -2.39e-05
horiz_align:centre_dist_init           -0.0004   9.27e-05     -4.015      0.000      -0.001      -0.000
horiz_align:whistdist_change            1.6343      0.344      4.751      0.000       0.960       2.308
centre_dist_change:dir_change           0.0003      0.000      2.834      0.005       0.000       0.001
centre_dist_change:centre_dist_init  1.285e-06   4.14e-07      3.102      0.002    4.73e-07     2.1e-06
centre_dist_change:whistdist           -0.0019      0.001     -2.031      0.042      -0.004   -6.63e-05
centre_dist_change:crossings        -5.761e-05   3.35e-05     -1.718      0.086      -0.000    8.13e-06
dir_change:dir_change2                  0.0724      0.048      1.495      0.135      -0.022       0.167
dir_change:log_amp_change              -0.1826      0.037     -4.905      0.000      -0.256      -0.110
dir_change:dist_1back                   0.0006      0.000      4.458      0.000       0.000       0.001
dir_change:centre_dist_init             0.0002      0.000      2.011      0.044    5.32e-06       0.000
dir_change:whistdist_change            -0.9302      0.329     -2.827      0.005      -1.575      -0.285
dir_change:crossings                   -0.0265      0.009     -2.871      0.004      -0.045      -0.008
dir_change2:log_amp_change             -0.1327      0.035     -3.831      0.000      -0.201      -0.065
dir_change2:dist_1back                  0.0003      0.000      1.977      0.048    2.63e-06       0.001
dir_change2:whistdist_change           -0.6522      0.293     -2.228      0.026      -1.226      -0.079
log_amp_change:dist_1back              -0.0005      0.000     -3.452      0.001      -0.001      -0.000
dist_1back:centre_dist_init          1.344e-06   3.54e-07      3.797      0.000     6.5e-07    2.04e-06
dist_1back:whistdist_change             0.0020      0.001      1.630      0.103      -0.000       0.004
dist_1back:crossings                  4.92e-05   3.36e-05      1.464      0.143   -1.67e-05       0.000
centre_dist_init:whistdist_change      -0.0018      0.001     -1.730      0.084      -0.004       0.000
centre_dist_init:crossings          -9.163e-05    3.7e-05     -2.479      0.013      -0.000   -1.92e-05
have_visited:crossings                  0.0264      0.012      2.207      0.027       0.003       0.050
whistdist:whistdist_change              2.9111      1.229      2.369      0.018       0.503       5.319
whistdist:crossings                    -0.2338      0.067     -3.477      0.001      -0.366      -0.102
t                                       0.0317      0.007      4.345      0.000       0.017       0.046
logdist:t                              -0.0071      0.001     -4.955      0.000      -0.010      -0.004
=======================================================================================================
\end{verbatim}
}
\bibliographystyle{unsrt}
\bibliography{pnas-sample}